\documentclass[amsmath,twocolumn,superscriptaddress,prb,aps]{revtex4}
\usepackage{amsthm,amsfonts,graphicx,verbatim, color}
\usepackage{graphicx}
\usepackage{bm}% bold math
\usepackage[utf8]{inputenc}
\let\oldmarginpar\marginpar
\renewcommand\marginpar[1]{\-\oldmarginpar[\raggedleft\footnotesize #1]%
{\raggedright\footnotesize #1}}

\newcommand{\be}{\begin{equation}}
\newcommand{\ee}{\end{equation}}
\newcommand{\bea}{\begin{eqnarray}}
\newcommand{\eea}{\end{eqnarray}}

\renewcommand{\epsilon}{\varepsilon}
\renewcommand{\vec}[1]{{\bf #1}}

\renewcommand{\cite}[1]{[\onlinecite{#1}]}

\newcommand{\sign}{\mathrm{sign}}

\begin{document}
\title{Rare region effects dominate weakly-disordered 3D Dirac points}
\author{Rahul Nandkishore}
 \affiliation{Princeton Center for Theoretical Science, Princeton University, Princeton, New Jersey 08544, USA}
\author{David A. Huse}
 \affiliation{Princeton Center for Theoretical Science, Princeton University, Princeton, New Jersey 08544, USA}
 \affiliation{Department of Physics, Princeton University, Princeton New Jersey 08544, USA}
\author{S. L. Sondhi}
 \affiliation{Department of Physics, Princeton University, Princeton New Jersey 08544, USA}

 \begin{abstract}
We study three-dimensional Dirac fermions with weak finite-range scalar potential disorder. In the clean system, the density of states vanishes quadratically at the Dirac point. Disorder is known to be perturbatively irrelevant, and previous theoretical work has assumed that the Dirac semimetal phase, characterized by a vanishing density of states, survives at weak disorder, with a finite disorder phase transition to a diffusive metal
with a non-vanishing density of states. In this paper we
%challenge this consensus. We
show that nonperturbative effects from rare regions, which are missed by conventional disorder-averaged calculations, instead give rise to a nonzero density of states for any nonzero disorder. Thus, there is no Dirac semimetal phase at non-zero disorder. The results are established both by a heuristic scaling argument and via a systematic saddle point analysis.
%Our analysis reveals that there is no regime of stability for the Dirac semimetal: the system has a non-zero (but exponentially small) density of states for arbitrarily weak disorder.
We also discuss transport near the Dirac point. At the Dirac point, we argue that transport is diffusive, and proceeds via hopping between rare resonances. As one moves in chemical potential away from the Dirac point, there are interesting intermediate-energy regimes
where the rare regions produce scattering resonances that determine
the DC conductivity. We derive a scaling theory of transport near disordered 3D Dirac points.  We also discuss the interplay of disorder with attractive interactions at the Dirac point, and the resulting granular superconducting and Bose glass phases.
Our results are relevant for all 3D systems with Dirac points, including Weyl semimetals.
\end{abstract}
\maketitle

\section{Introduction}
The discovery of two-dimensional (2D) Dirac systems such as graphene and the surface states of topological insulators has sparked an explosion of activity in condensed matter physics \cite{CastroRMP, KaneHasan}. Such materials, which are gapped everywhere except at isolated points in the Brillouin zone, play host to an abundance of new physics. In particular, when the chemical potential is placed at the `Dirac point', they display behavior that is intermediate between metals and insulators, in that the spectrum is gapless, but displays a vanishing low-energy density of states (DOS). The theoretical prediction \cite{Murakami, Wan, Balents, Turner, Zaheer} and experimental discovery \cite{expt1, expt2, expt3, expt4} of three dimensional (3D) Dirac points provides a higher dimensional version of this behavior, and has ignited a blaze of interest in 3D Dirac points.

When the Fermi level lies precisely at the Dirac point of a clean 3D system, the density of states vanishes and the mean free path diverges. The consensus in the theory literature, from original work by Fradkin \cite{Fradkin} in the 1980's to more recent work on Weyl semimetals \cite{Goswami, Hosur, Garate, MoreWeyl1, MoreWeyl2, MoreWeyl3, MoreWeyl4}, is that weak disorder is perturbatively irrelevant at 3D Dirac points, so that sufficiently weak disorder does not affect the vanishing of the density of states (DOS) at the Dirac point, or the divergence of the mean free path. Thus it has been assumed that there is a `Dirac semimetal' phase (characterized by a vanishing DOS), which survives at weak disorder, and which undergoes a quantum phase transition at a critical disorder strength to a diffusive metal.

In this paper we show that this longstanding theoretical consensus is inaccurate, and that the Dirac semimetal phase does not exist at non-zero disorder. The source of the inaccuracy is non-perturbative rare region effects, which can dominate the %weak disorder
physics at particle-hole symmetric points \cite{DiracSC, Lesik}, and which were ignored in all previous analyses. When these rare region effects are correctly accounted for, the density of states at the Dirac point remains non-zero even for arbitrarily weak disorder, and the mean free path remains finite. There is no quantum phase transition at finite disorder. Rather, the Dirac semimetal only exists in the limit of vanishing disorder strength.

There is some similarity between the rare region effects discussed in this paper and the phenomenon of Lifshitz tails \cite{Lifshitz1, Lifshitz2, Lifshitz3, Lifshitz4}, of which a remarkably clear exposition can be found in \cite{Yaida}. However, there are also important differences. Whereas Lifshitz tails involve exponentially localized states \cite{PWA} which exist inside a band gap, in the problem of interest to us there is no band gap, and thus a straightforward mapping to the Lifshitz tail problem is clearly impossible. The origin of the non-zero DOS in the present problem is more subtle, with the non-zero DOS arising due to power law bound resonances which co-exist with an extended continuum. The spinor nature of the wave function is also an essential requirement for the rare region effects we analyze, which are thus particular to Dirac fermion systems.

Our paper is organized as follows: in Section II we introduce the model of interest to us, and explain why previous theoretical analyses have (plausibly but erroneously) concluded that weak disorder can be ignored. In Section III we provide a heuristic scaling argument that suggests that a nonzero density of states arises for any nonzero disorder. In Section IV, we re-derive this result by means of a systematic saddle point analysis, conclusively establishing that the density of states is nonzero for any nonzero disorder. In Section V, we examine transport near the Dirac point, and discuss the multiple distinct transport regimes that arise as we tune the chemical potential away from the Dirac point. In Section VI we  discuss the interplay of disorder and interactions, paying particular attention to the granular superconductor and Bose glass phases that may arise. We summarize our results in Section VII. The Appendix derives some results on massless three dimensional Dirac equations in spherically symmetric potentials. These results are used extensively in the main text.

\section{Model and background}
The low energy Hamiltonian of interest takes the form
\begin{equation}
H = \sum_{a = 1}^{2N} \bigg[\int d^3\vec{k} v^a \psi^{\dag}_{a}(\vec{k})\vec{\sigma}\cdot\vec{k} \psi_{a}(\vec{k}) +
\int d^3\vec{r}V(\vec{r}) \psi^{\dag}_{a}(\vec{r})  \psi_{a}(\vec{r})\bigg] ,
\end{equation}
where the two-component spinor $\psi_a(\vec{k})$ represents a state near the Dirac point $a$, with a momentum $\vec{k}$ relative to the Dirac point, $\psi_a(\vec{r})$ is its Fourier transform,
and $V(\vec{r})$ is a random scalar potential which is short range correlated and has mean zero. In general in a condensed matter system the dispersion about each Dirac point would be anisotropic, but for simplicity we consider the isotropic case.

In any lattice model with emergent Dirac fermions, Dirac points always come in pairs. For simplicity, we put a UV cutoff on the random potential so that it
%However, the random scalar potential (unlike a random mass)
does not produce scattering between Dirac points, thus the different Dirac points are all decoupled. It is therefore sufficient for us to consider a single Dirac point with random scalar potential disorder, i.e.
\begin{equation}
H =  \int d^3\vec{k} v \psi^{\dag}  \vec{\sigma}\cdot  \vec{k} \psi +\int d^3\vec{r}V(\vec{r}) \psi^{\dag}  \psi   ~.
\end{equation}
This model most clearly exposes the relevant physics. Additional Dirac points can be retained in the analysis without changing the essential results. Near the Dirac point, in the clean limit $V = 0$, the low energy DOS (per Dirac point per unit volume) then vanishes as $\nu(E) \approx \frac{E^2}{2 \pi^2 (\hbar v)^3}$.  %For convenience, we will often work in units where $\hbar=v=1$.

We now add weak quenched scalar potential disorder (strong disorder has been studied in \cite{Arovas}). A simple and highly intuitive argument for the irrelevance of scalar potential disorder proceeds as follows. An energy scale $E$ sets a length scale $\hbar v / E$.  Assuming short range correlated disorder with $\langle V(\vec{r}) \rangle = 0$ and $\langle V(\vec{R})V(\vec{R}+\vec{r})\rangle = \mu_0^2f(r/b)$, with $f(0)=1$ and $f(x)$ decaying exponentially
% If I remember correctly then power law correlated disorder has a different Harris counting
 for $x>1$, and averaging the disorder over a volume $(\hbar v/E)^3$ using the central limit theorem, we conclude that the magnitude of the average potential over a length scale $\hbar v/E$ will be $|\delta V|\sim  \mu_0 E^{3/2}$. The ratio $|\delta V|/E \sim \mu_0 E^{1/2}$ vanishes as $E \rightarrow 0$, so one might conclude that in the asymptotic zero energy limit, the typical average potential vanishes more rapidly than the energy itself, and can thus be ignored.

An alternative argument for the perturbative irrelevance of disorder proceeds \cite{Abrikosov-Gorkov, Hosur} by evaluating the electron self energy $\Sigma$, which yields
$ \Sigma(\omega, \vec{k}\rightarrow 0) \sim \mu_0^2 b^3 \omega^2$.
This vanishes more rapidly than $\omega$ at low energies and thus allows existence of sharp quasiparticles.
Similarly, a self consistent Born approximation (SCBA) for the mean free path $l$ leads to % the self consistency condition
\begin{equation}
\frac{\hbar v}{l} = \frac{\hbar v}{l} \mu_0^2 b^3 \int_0^{\Lambda} \frac{\nu(E) dE}{E^2 + \hbar^2v^2/l^2}  \label{eq: SCBA}
\end{equation}
%
%when the chemical potential is
at the Dirac point; $\Lambda$ is a UV cutoff and $\nu(E) \sim E^2$.  For sufficiently weak disorder $\mu_0 \rightarrow 0$, this admits only the trivial solution $1/l = 0$ [at energies away from the Dirac point, $l$ diverges as $l \sim 1/(\mu_0^2 b^3 E^2)$ within SCBA]. This is in sharp contrast to two-dimensional Dirac materials, where, within SCBA, disorder produces a crossover to diffusive behavior at long length scales \cite{Ando}. The difference arises because the DOS vanishes more rapidly in $3D$, making disorder perturbatively irrelevant instead of marginal.

We note that in the above equation, we introduced a UV cutoff on the Dirac equation $\Lambda$. This in turn defines a length scale $a = \hbar v / \Lambda$, which should be of order the lattice scale. We henceforth set $\hbar = v = a=1$ for convenience.  All lengths are measured in units of $a$, all energies are measured in units of
$\hbar v/a$ and all times are measured in units of $a/v$. %i.e. t
This consistent set of natural units will be used throughout this paper, although on occasion we will choose to display the factors of $\hbar, v$ and $a$ explicitly.

Finally, a simple RG analysis also suggests that scalar potential disorder is irrelevant. The argument proceeds as follows: working with the Matsubara field integral
and ensemble averaging over Guassian-distributed disorder using the replica trick (see e.g. \cite{DiracSC}) gives rise to a quartic term of the form
$\sim\mu_0^2 \int d\tau d \tau' d^3x \bar \psi(\vec{x},\tau) \psi(\vec{x},\tau) \psi(\vec{x},\tau') \psi(\vec{x},\tau') $. %, where $W$ is the disorder strength.
Straightforward power counting then reveals that $\mu_0$ is irrelevant in the renormalization group sense at tree level.

As a result of all these excellent and intuitive arguments for the irrelevance of disorder, it has long been believed that there exists a Dirac semimetal phase characterized by a vanishing density of states and a diverging mean free path, which survives at weak disorder, with a quantum phase transition to a diffusive metal occurring at a nonzero critical disorder strength. It is the objective of this paper to establish that this belief is incorrect - there is no Dirac semimetal phase at non-zero disorder. Rather the system has a non-vanishing density of states and a finite mean free path for any nonzero disorder. This new result arises due to the effect of exponentially rare regions which host critically localized resonances, and which are missed by conventional disorder-averaged calculations. An analysis of these rare regions effects will occupy the majority of this paper.

%we will show in this paper that even arbitrarily weak disorder cannot be neglected, and in fact dominates the physics at the Dirac point.

\section{A heuristic argument for the importance of disorder}
In this section, we provide a heuristic argument for the importance of disorder. To this end, we will invoke a somewhat artificial model of disorder, which nevertheless captures the essential physics (a more realistic treatment of disorder will be provided in the following section). The model of disorder we consider is one where we introduce Poisson distributed `impurities', with a mean density of impurities $n$. Each impurity consists of a spherically symmetric scalar potential of the form $V(r) = \lambda \Theta(b - r) + \lambda \epsilon(r) \Theta(r-b)$, where $b$ is a fixed length scale, $\lambda$ (the `strength' of the impurity) is taken from a Gaussian distribution $P(\lambda)$ with mean zero and variance $\mu_0^2/(nb^3)$, and $\epsilon$ obeys $\epsilon(b) = 1$ and falls off at least as fast as $1/r^4$ at long distances. i.e. each impurity is modeled as a scalar potential well with a `tail' that falls off at least as fast as $1/r^4$. The precise form of the tail will not be important for our argument. We work in the weak disorder limit, which corresponds to $\mu_0 \ll 1/b$ and $\mu_0^2 \ll nb$.  For convenience, we now take $nb^3=1$, although our results are readily generalized to other cases with no important changes.

It can be shown (see Appendix) that a single impurity will trap a bound state for specific values of $\lambda b$. The bound states take the form
\begin{equation}
\psi^{\pm} = f(r) \phi^{\pm}_{j, j_z} +i  g(r) \phi^{\mp}_{j, j_z}; \label{generalsoln}
\end{equation}
where $f$ and $g$ are purely radial functions with no angular dependence, and the $\phi^{\pm}$ are two component spinors (detailed in the Appendix) which have definite total angular momentum $j$ and which have orbital angular momentum differing by one. Bound states arise with all values of $j$. However, bound states with high $j$ require a deeper or wider well than bound states with smaller $j$.  In the weak disorder limit, the physics near the Dirac point is dominated by bound states on rare very strong impurities with $j=1/2$ (which are linear superpositions of states with orbital angular momentum zero and one). We emphasize that for the $j=1/2$ solutions, $|\psi|^2$ is isotropic i.e. the $j=1/2$ bound state solutions have an isotropic probability density.

The existence of bound states is not sensitive to the form of $\epsilon(r)$, and it is convenient to take an $\epsilon(r)$ that falls off infinitely fast (corresponding to a square well). In this case the bound states arise at $\lambda = \lambda_c \approx m \pi/b$, for nonzero integer $m$, and have $f(r\rightarrow \infty) \sim 1/r^2$ and $g(r\rightarrow \infty) =0$. The position of $\lambda_c$ shifts if we make a different choice for $\epsilon(r)$, and the subleading piece $g(r)$ changes (e.g. $g(r) \sim 1/r^5$ for $\epsilon(r) \sim 1/r^4$) but the existence of bound states and the leading $1/r^2$ falloff of the wave function do not change, and the scaling argument that we will now present is unaltered. We therefore stick to the `square well' potential, which has bound states for $\lambda b \approx m \pi$. Recall that $P(\lambda) \sim \exp(-\lambda^2 / 2\mu_0^2)$.  In the weak disorder limit the physics is dominated by the $j=1/2$ bound states, which have $m= \pm1$ and  probability density $P(\lambda_c) \sim \exp(-\tilde C \pi^2 / (2\mu_0^2b^2))$, where $\tilde C$ is a numerical prefactor of order one. % (which depends on whether we are working with a square well or a well  $\epsilon(r)$.

Now, the bound states in question are rather delicate. Unlike the case of Lifshitz tails, where the  bound states are exponentially bound, here the states are only
power law bound. Moreover (also unlike the case of Lifshitz tails), true bound states arise only for precise values of $\lambda = \lambda_c$. %To make further
 %Our argument relies on certain properties of the 3D Dirac equation in a spherically symmetric potential, which are worked out in the appendix.
In an infinite sample, even an infinitesimal deviation from $\lambda = \lambda_c$ or $E =0$ leads to the disappearance of the bound state (indeed the solution becomes
non-normalizable). However, the disappearance of the bound state becomes apparent only at a very large length scale $R$, which diverges as
$\lambda \rightarrow \lambda_c$ and $E \rightarrow 0$. On length scales less than $R$, the solution for $\lambda $ close to but not quite $\lambda_c$ is
indistinguishable from the truly bound state. In a system with a nonzero density of impurities hosting almost bound states, where $R$ exceeds
the spacing between such almost bound states, such small deviations from $\lambda_c$ should not matter. Thus, we expect that the density of
quasi-bound states will be
\begin{equation}
n_{bound} = n P(\lambda_c) \delta \lambda \sim n \exp \left(- \frac{\tilde C \pi^2}{2\mu_0^2b^2} \right) \delta \lambda
\end{equation}
where $\delta \lambda$ is a to-be-determined quantity that tells us how close we have to get to $\lambda_c$ in order to have a state that looks effectively bound.
Moreover, these quasi-bound states will not all be strictly at zero energy. Rather they will be spread over an energy window of width $\delta E$. The contribution
to the density of states coming from quasi bound states will then take the form
\begin{equation}
\nu(E) \sim n \exp \left(- \frac{\tilde C \pi^2}{2\mu_0^2b^2} \right) \frac{\delta \lambda }{\delta E} = \nu_0 \frac{\delta \lambda }{ \delta E} \label{dos1'}
\end{equation}
where we have defined
\begin{equation}
\nu_0 = n \exp \left(- \frac{\tilde C \pi^2}{2\mu_0^2b^2} \right) \label{dos1}
\end{equation}

To make further progress requires understanding what happens when $\lambda$ and $E$ are slightly perturbed from $\lambda_c$ and $0$ respectively. To this end, it is instructive to calculate the scattering cross section $\sigma(\lambda, E)$ of a single well. The scattering cross section $\sigma$ is given by the formula

\begin{equation}
\sigma = \frac{4\pi}{k^2} \sin^2 \delta = \frac{4 \pi }{E^2} \sin^2 \delta,
\end{equation}
 where $\delta$ is the phase shift \cite{Perkins}. We now have to determine the phase shift $\delta$.
 Since we have already established that the nature of the `tail' of $\epsilon(r)$ does not qualitatively alter the physics, we model the well as being simply a square well - this greatly simplifies the calculation.

In the absence of a scattering potential, the Dirac equation (in polar co-ordinates) has a solution which is a spherical Bessel function of the first kind, which at long distances has the asymptotic form $J_{\alpha}(kr) \sim \frac{1}{\sqrt{kr}} \cos(kr - \alpha \pi/2 - \pi/4)$. In the presence of a scattering potential, the solution (see Appendix) is $\psi \sim \big(A' J_{\alpha}(kr) + B' K_{\alpha} (kr) \big) \sim A' \cos(kr - \alpha \pi/2 - \pi/4) + B' \sin(kr - \alpha \pi/2 - \pi/4) \sim C \cos (kr - \alpha \pi/2 - \pi/4 - \delta)$, where $\delta$ is the phase shift. Application of standard trigonometric identities, as well as the results from the Appendix for $A'$ and $B'$ (see also \cite{Dombey}), then leads to the result

\begin{widetext}
\begin{equation}
\tan \delta = \frac{\sign \big(\frac{E}{E-v}\big)J_{3/2}(|E|b) J_{1/2}(|E-V| b) - J_{1/2}(|E| b) J_{3/2}(|E-V| b )}{\sign \big(\frac{E}{E-V}\big) J_{1/2}(|E-V| b) K_{3/2}(|E| b) - J_{3/2}(|E-V| b) K_{1/2} (|E| b)}\label{phase shift}
\end{equation}
\end{widetext}
This equation contains a great deal of physics. The magical values of $\lambda = \lambda_c$ which give rise to bound states are revealed as {\it resonances}, which correspond to phase shifts $\delta = \pi/2$. These resonances `pull' some density of states out of the continuum and down to zero energy (in the form of bound states). Meanwhile, in the scaling limit $E \rightarrow 0$, the cross section is a tightly peaked Lorentzian, with
\begin{eqnarray}
\sigma(E, \lambda) &\sim& \frac{ E^2 b^2 }{(\lambda - \lambda_c(E))^2  + E^4 b^2}; \nonumber\\  \lambda_c(E) - \lambda_c(0) &\sim& E; \qquad \lambda_c(0) \sim \pm \pi / b
\end{eqnarray}
From this we conclude that there is a line of resonances in the $\lambda, E$ plane, and that these resonances have width $\sim b E^2$ in both $\lambda$ and $E$. This leads us to the scaling $\delta \lambda \sim \delta E \sim b E^2$. Substituting this into (\ref{dos1'}) tells us that the low energy density of states is just $\nu_0$, given by (\ref{dos1}).

When making this estimate, we have not taken the non-resonant `extended' states into account, thus this estimate is valid only on scales $E < \sqrt{\nu_0}$, where the density of states $\nu_0$ from special wells exceeds the DOS $\sim E^2$ from the extended states. However, in this regime, our scaling theory reveals that the DOS is given by (\ref{dos1}), and is non-zero for arbitrarily weak disorder. This establishes that even though disorder is perturbatively irrelevant at the Dirac point, it cannot be neglected.

We close this section by highlighting one important point. The white noise limit $b \rightarrow 0$ is `pathological'. In the limit $b \rightarrow 0$
at any, even very small, fixed $\mu_0^2 b^3$ (necessary to keep the disorder strength constant), the system can not remain in the fully weak
disorder limit $\mu_0\ll 1/b$ and these `rare quasi bound states' become not at all rare.
%, and we cease to obtain sensible results. The pathological nature
This aspect of the white noise limit will become clear in the saddle point calculation presented in the following section.

\section{A systematic calculation of the density of states}
In the previous section we provided a heuristic scaling argument suggesting that the DOS is non-zero in the presence of arbitrarily weak disorder. We now rederive this result using `standard' techniques. We follow the route taken in \cite{Yaida}, suitably generalized to the present problem.
% Then, I show (again following the discussion in \cite{Yaida}) how the same results can be obtained using the supersymmetry technique.
We begin by noting that we are dealing with a system governed by the Dirac Hamiltonian
\begin{equation}
\left[-i \sigma_{i} \partial_{i} + V(\vec{x}) {\bf 1} \right] \psi^V_n(\vec{x}) = E_n^V \psi^V_n(\vec{x})
\end{equation}
where $\sigma_{i}$ is a Pauli matrix, and the $\psi^V_n$ are the two-component spinor eigenfunctions which satisfy the above equation for eigenenergies
$E^V_n$ in a given random scalar potential $V$.  Repeated indices are summed over. The density of states per unit volume at an energy $E$ and for a given
disorder configuration $V$,  $\nu^V(E)$, can be expressed as
\begin{equation}
\nu^V = \frac{1}{L^3}\sum_n \delta(E - E^V_n) \label{dos2}
\end{equation}
where $L$ is the linear size of the system. We now introduce a spinor Lagrange multiplier field $\chi$ and a scalar Lagrange multiplier $\Upsilon$ to rewrite this as
\begin{widetext}
\begin{equation}
\nu^V = \frac{1}{L^3} \int D[\psi(\vec{x}), \chi(\vec{x}), \Upsilon] \exp \bigg[i \int  d^3 x \chi^{\dag}(\vec{x}) \bigg( E + i \sigma_{i} \partial_{i} - V(\vec{x}) \bigg) \psi(\vec{x}) + i \Upsilon \bigg( \big[\int d^3x \psi^{\dag}(\vec{x})\psi(\vec{x})\big] -1  \bigg) \bigg]
\end{equation}
\end{widetext}
Integrating out the Lagrange multiplier fields gives us a delta function which picks out only those configurations $\psi(\vec{x})$ which satisfy the Dirac equation with $E = E_n^V$ and are properly normalized. The functional integral over $\psi(\vec{x})$ then reproduces (\ref{dos2}). Note that in order for the exponent to be properly dimensionless when $\psi$ is properly normalized,
%has dimensions of probability density $[\psi ] = L^{-3/2}$,
the scalar $\Upsilon$ must be a pure number, whereas $\chi$ must have dimensions of $[\psi ]/[E]$.
%1/energy * probability amplitude.

So far we have discussed the DOS for a specific disorder realization. We now average over disorder (assuming that the disorder is a Gaussian random variable, which is short range correlated with a correlation length $\xi$), to obtain a disorder averaged density of states $\overline{\nu}$, which takes the form
\begin{equation}
\overline{\nu} = \frac{1}{L^3} \int D[V, \psi, \tilde \chi, \tilde \Upsilon] \exp \bigg[- S\bigg], \label{eq: dosformal}
\end{equation}
 where
\begin{widetext}
\begin{eqnarray}
%\overline{\nu} &=& \frac{1}{L^d} \int D[V, \psi, \tilde \lambda, \tilde \mu_0] \exp \bigg[- S\bigg] \\
S &=& \frac{1}{2W^2} \int d^3 x d^3x' V(\vec{x}) V(\vec{x'})K^{-1}(\vec{x}-\vec{x'}) - \frac{1}{W} \int d^3x    \tilde \chi^{\dag}(\vec{x}) \bigg( E + i  \sigma_{i} \partial_{i} - V(\vec{x}) \bigg) \psi(\vec{x}) \label{action} \nonumber\\&+& \tilde \Upsilon \bigg( \big[\int d^3x \psi^{\dag}(\vec{x})\psi(\vec{x})\big] -1  \bigg) \bigg] \nonumber\\
\end{eqnarray}
\end{widetext}
We have performed one formal manipulation, defining the rescaled variables $\tilde \chi^{\dag} = i \chi^{\dag}W $ and $\tilde \Upsilon = i \Upsilon$.
We have scaled the lengths by the microscopic length scale $a$ (which has been set equal to one), so $\tilde \chi^{\dag} $ and $\psi$ are now dimensionless.  $W$ measures the disorder strength,
with $W^2 \sim \mu_0^2\xi^3$.
%now has dimensions of a probability density, whereas $\Upsilon$ is now a scalar. The scale $a$ is undetermined, but should be of order e.g. the lattice scale.
Meanwhile, $K$ is the correlation function for the disorder,
%(which is a delta function in the white noise limit),
and we have defined $K^{-1}$ according to $\int d^3y K^{-1}(\vec{y}-\vec{y'}) K(\vec{y}-\vec{y''}) = \delta^3(\vec{y'}-\vec{y''})$.
%However, (as we shall see) the white noise limit is pathological, and some UV regulation is necessary to get sensible results.
We assume that K is an isotropic and normalized function which is short ranged with a characteristic scale $\xi$ (for definiteness, we could take $K$ to be a normalized isotropic Gaussian with width $\xi$, but the results will be independent of the precise shape of $K$).

We now make the one essential approximation required by our approach: we calculate the density of states $\overline{\nu}$ in a saddle point approximation. The saddle point equations obtained by varying $V$, $\tilde \chi$, $\tilde \Upsilon$, $\psi^{\dag}$, and $\psi$ respectively are:
\begin{eqnarray}
- W  \int d^3x' K(\vec{x}-\vec{x'}) \tilde \chi^{\dag}(\vec{x'}) \psi(\vec{x'}) &=& V(\vec{x}) \label{potential'}\\
\left[-i  \sigma_{i} \partial_{i} + V(\vec{x}) {\bf 1} \right] \psi(\vec{x}) &=& E \psi(\vec{x}) \label{psi}\\
\int d^3 x \psi^{\dag}(\vec{x}) \psi(\vec{x}) &=& 1\\
\Upsilon \psi(\vec{x}) &=& 0 \\
\tilde \chi^{\dag} \left[E +i \sigma_{i} \partial_{i} - V(\vec{x}) {\bf 1} \right] &=& 0. \label{lambda}
\end{eqnarray}
 Now, we note that the equation for $\tilde \chi^{\dag}$, (\ref{lambda}), is just the Hermitian conjugate of the equation for $\psi$, (\ref{psi}). Thus, we take $\tilde \chi^{\dag} = \chi_0 \psi^{\dag}$, where $\chi_0$ is a scalar. We want to search for solutions at real energies, so we want the saddle point Hamiltonian to be Hermitian. This then demands that we should take $\chi_0$ to be a real number, although it could be either positive or negative. Substituting into (\ref{potential'}) tells us that within the saddle point approximation, the DOS is dominated by potential configurations with
\begin{equation}
V(\vec{x}) = - \chi_0 W  \int d^3x' K(\vec{x}-\vec{x'})  \psi^{\dag}(\vec{x'}) \psi(\vec{x'}).\nonumber
\end{equation}
Thus we find that the equations of motion all boil down to a single (non-linear) integro-differential equation which takes the form
\begin{equation}
\left[-i \sigma_{i} \partial_{i} -  \chi_0  W  \int K (\vec{x}-\vec{x'}) \psi^{\dag}(\vec{x'}) \psi(\vec{x'}) {\bf 1} \right] \psi(\vec{x}) = 0; \label{nonlineareqn}
\end{equation}
where we have specialized to $E=0$ and require that the solutions $\psi$ be properly normalized $\int d^3 x |\psi(\vec{x})|^2 = 1$. We emphasize that we are allowed to tune $\chi_0$ in order to find a solution. The contribution of a particular solution to the disorder averaged density of states is found by substituting the saddle point solution into (\ref{eq: dosformal}). This yields
\begin{widetext}
\begin{equation}
\delta \overline{\nu} = \frac{1}{L^3} \exp \left(- \frac{\chi_0^2}{2} \int d^3x d^3x' \psi^{\dag}(\vec{x}) \psi(\vec{x}) K(\vec{x}-\vec{x'}) \psi^{\dag}(\vec{x'}) \psi(\vec{x'})\right) \label{nonwhitenoisesaddle}
\end{equation}
\end{widetext}
We must sum over all saddle point solutions to accurately obtain the density of states.

We note that for any solution $\psi_0$ of the above equation with $E, \chi_0$, there will be a corresponding solution to the same equation with $E\rightarrow -E, \chi_0 \rightarrow -\chi_0, \psi_0 \rightarrow C \psi_0$, where $C$ is the particle hole symmetry operator, and this corresponding solution will have the same cost action. Thus, the DOS will be even in energy. It is sufficient for our present purposes to determine the DOS at $E=0$.

\subsection{SCBA}
The equation (\ref{nonlineareqn}) clearly has saddle point solutions that are plane waves $\sim \frac{1}{L^{3/2}} e^{i\vec{k}\cdot \vec{x}}$, with $k = |E|$. Substituting into (\ref{nonwhitenoisesaddle}) tells us that the contribution of the plane wave saddle points to the density of states is
\begin{equation}
\delta \overline{\nu}(E) \sim \frac{1}{L^3}\sum_{\vec{k}} \exp\left(- \frac{\chi_0^2 }{2L^3}\right) \delta(E - v k) \sim \frac{1}{L^3}\sum_{\vec{k}} \delta(E - v k)
\end{equation}
Thus, the density of states is just equal to the number of plane wave solutions at a given energy, divided by the volume. In the $E \rightarrow 0$ limit, the number of plane wave solutions in a window of energies between $E$ and $E + \delta E$ scales as $L^3 E^2 \delta E$, thus, the density of states coming from these `plane wave' saddle points scales as $E^2$, and vanishes at zero energy. The various arguments for the perturbative irrelevance of disorder outlined in Section II essentially amount to the statement that perturbation theory about these translation invariant saddle points converges. However, we will now proceed to show that there are additional solutions which satisfy the saddle point equations, and while these other solutions have a weight that is exponentially small in weak disorder, the density of these other saddle points does not vanish as $E \rightarrow 0$. It is these other saddle points (which correspond to rare resonances) this will give rise to a non-zero density of states at $E=0$.

\subsection{A saddle point treatment of rare regions}
Motivated by the scaling analysis in the previous section, we look for localized and normalizable solutions to the saddle point equations. We assume for convenience that the localized solution is centered at the origin. It is convenient to interpret (\ref{nonlineareqn}) as a Dirac equation in an effective potential
\begin{equation}
V^{eff}(\vec{x}) = - \chi_0 W \int d^3 x' K(\vec{x} - \vec{x'}) \psi^{\dag}(\vec{x}') \psi(\vec{x'}) \label{potential}
\end{equation}
Guided by our previous analysis, we expect to find a solution of the form (\ref{generalsoln}) with total orbital angular momentum $j=1/2$, and such a solution has the property that $|\psi|^2$ is isotropic. If $K$ is an isotropic function, it then follows that the effective potential is spherically symmetric. Thus, (\ref{nonlineareqn}) can be re-interpreted as a Dirac equation in a spherically symmetric potential, where the shape of the potential must be determined self consistently, and where the potential profile tracks the probability density for a bound state wave function with total angular momentum $j=1/2$. (There will also be solutions corresponding to higher values of $j$, but these will require a deeper or wider self consistent potential well and thus will have exponentially suppressed contribution to the DOS, such that the DOS will be dominated by solutions with $j=1/2$).

Now, we have already assumed that the kernel $K$ is sharply peaked e.g. $K(r) \sim \frac{1}{\xi^3} \exp(-r^2/\xi^2)$. On long length scales $r \gg \xi$, $K$ can be modeled as a delta function, leading to the simpler equation
\begin{equation}
\left[-i \partial_{i} \partial_{i} -  \chi_0  W \psi^{\dag}(\vec{x}) \psi(\vec{x}) {\bf 1} \right] \psi(\vec{x}) = 0; \label{whitenoise}
\end{equation}

Making the ansatz (\ref{generalsoln}) with $j=1/2$, this can be written out in components as
\begin{eqnarray}
V ( | f|^2 + | g|^2) g &=& (\partial_{ r} + \frac{2}{ r})  f\\
- V( | f|^2 + | g|^2)  f &=& \partial_{r}  g
\end{eqnarray}
where we have introduced the shorthand variable $X = \chi_0 W/4\pi$. We now imagine constructing a power series solution $f = f_1 + f_2 + f_3...$ and $g = g_1 + g_2 + g_3+...$, where each successive term is higher order in $1/r$. Guided by our earlier work on the linear Dirac equation (see Appendix), we look for a solution where $f \sim 1/r^2$ and $g$ falls off faster. This leads to a long distance solution that has the form
\begin{equation}
f \sim \frac{A}{r^2}  - \frac{X A^5}{6 r^8} + O(\frac{X^4 A^9}{r^{14}})\quad g \sim \frac{X A^3}{5 r^5} - \frac{23 X^3 A^7}{550 r^{11}}+ O(\frac{X^5 A^{11}}{r^{17}})
\end{equation}
with an undetermined scale factor $A$. We can readily see that this takes the form of a perturbation series in the small parameter $X^2 A^4 / r^6 \ll 1$. We now check for self consistency. We note that the solution identified above describes an effective potential that falls off as $1/r^4$ at large $r$. We have already identified (see Appendix) that the Dirac equation in a spherically symmetric potential that falls off as $1/r^4$ has a solution where $f \sim 1/r^2$ and $g \sim 1/r^5$. Thus, the solution we have constructed is indeed a correct self consistent solution of the non-linear integro-differential equation (\ref{nonlineareqn}) at large distances.

The expansion introduced above breaks down at $r_c = X^{1/3} A^{2/3} = ( \chi_0 W A^2 /4\pi)^{1/3}$. A numerical investigation of the equation (\ref{whitenoise}) reveals that solutions with the asymptotics identified above are generally singular at $r=0$. However, at small distances $r \le \xi$, modeling the disorder correlation function $K$ as a delta function  is clearly inappropriate, and thus we cannot work with (\ref{whitenoise}), but instead we must work with the full integro-differential equation (\ref{nonlineareqn}). On distances $r < \xi$, the convolution with $K$ produces an effective potential which is roughly constant.

This then implies that we are solving the Dirac equation in a spherically symmetric well that is of constant depth $\sim \chi_0 W $ on length scales less than $\xi$, but has a $1/r^4$ tail. We know what the solutions to this problem look like (from the Appendix): they have   $|\psi| \approx A$ at short distances $r<\xi$  and $|\psi| \approx A \xi^2 /r^2$ at long distances $r>\xi$. Normalization fixes $A^2 \approx \xi^{-3}$. The self consistent potential defined by (\ref{potential}) then is uniform with depth $\chi_0 W A^2 = \chi_0 W \xi^{-3}$ at short distances $r<\xi$ and falls off as $\chi_0 W  \xi / r^4$ at long distances $r>\xi$. We recall that we are free to tune $\chi_0$ to find a solution. Given the results derived in the Appendix, we expect that bound states will exist for an infinite discrete set of $\chi_0$. However, since the contribution to the DOS falls off exponentially with $\chi_0$ (Eq. \ref{nonwhitenoisesaddle}), the dominant contribution will come from the smallest value of $\chi_0$ that allows us to have a solution. The smallest value of $\chi_0$ that allows for a solution has  $\chi_0 W A^2 \xi \approx 4$ (see Appendix, Fig. 2) i.e. $\chi_0 \sim 1/(W \xi A^2)$.

%Now for $\lambda_0 \sim 1/(W \xi A^2)$ the asymptotic long distance expansion for the solution breaks down at $r_c = ( \lambda_0 W A^2 /4\pi)^{1/3}$.

Thus we have shown that there exists a normalizable and localized solution to the saddle point equations where the wave function falls off as $A/r^2$ at large distances, with $A^2 \sim \xi^{-3}$ and $\chi_0 \sim \frac{1}{W \xi A^2}$. Substituting this into the expression for the DOS (\ref{nonwhitenoisesaddle}) and approximating $\int d^3x d^3x' \psi^{\dag}(\vec{x}) \psi(\vec{x}) K(\vec{x}-\vec{x'}) \psi^{\dag}(\vec{x'})\psi(\vec{x'})  \approx \int d^3 x |\psi|^4$ we find that the saddle point solution identified above makes a contribution to the DOS of order
\begin{equation}
\delta \overline{\nu} \sim \frac{1}{L^3} \exp \left(- C \frac{\xi}{2W^2}  \right) \sim  \frac{1}{L^3} \exp \left(- C \frac{\xi (\hbar v)^2}{2W^2 a^3}  \right)
\end{equation}
where in the final expression we have restored $\hbar, v$ and $a$ for clarity. Here C is a numerical pre factor (expected to be of order one), which cannot be determined without actually solving the full integro-differential equation (as opposed to showing a solution exists and identifying its asymptotics).

We note that thus far we have evaluated the contribution to the disorder averaged DOS from a single localized saddle point solution centered at the origin. However, the localized saddle point solution could be centered anywhere in the sample and we must sum over all possible locations of the bound state centre i.e. there are an extensive number of saddle points of this form contributing to the density of states. The summation over the centre of mass co-ordinates cancels the $1/L^3$ factor in the above equation. Another way to state this result is to note that while the {\it density} of plane wave solutions to the saddle point equations vanishes as $E \rightarrow 0$, the density of localized solutions to the saddle point equations does not vanish as $E \rightarrow 0$, so that the localized solutions to the saddle point equations actually dominate the low energy DOS. This we obtain a final expression for the contribution to the DOS from localized solutions to the saddle point equation that takes the form
\begin{equation}
\delta \overline{\nu}(E=0) \sim  \exp \left(- C \frac{\xi }{2W^2}  \right)  \label{soln}
\end{equation}
Determining the precise constant $C$ in the exponential requires determining the precise shape of the localized solution everywhere (i.e. not just the asymptotics), whereas determining the pre-exponential factor requires a consideration of fluctuations about the saddle point (for more details on this procedure, see \cite{Yaida}). We defer consideration of these issues to future work. However, we note that the scaling of the DOS with disorder strength and well radius ($\exp(-\xi/W^2)$ ) is the same as that from the heuristic scaling approach employed in Section III, if we identify $W $ with the rms potential in the well $W^2a^3 \sim \mu_0^2 b^3$, and if we identify $b \sim \xi$. Thus, a systematic saddle point calculation reveals that the heuristic scaling approach developed in Section III obtains essentially correct results. It also reveals the flaw in the perturbative arguments detailed in Section II: those arguments only consider fluctuations about the wrong (i.e. translation invariant) saddle point, whereas the physics is dominated by different, translation symmetry breaking, localized saddle points.

We note that we have only taken into account the localized solutions to (\ref{nonlineareqn}) which minimize the cost action (the solutions with $j=1/2$). There will be additional localized solutions with higher total angular momentum, but these will have a larger cost action, and hence will make an exponentially smaller contribution to the DOS. Still, given the likely existence of higher angular momentum saddle points, (\ref{soln}) should properly be viewed as a lower bound on the DOS.

We note that in the white noise limit $\xi \rightarrow 0$, the contribution from these localized solutions become of order one. However, the white noise limit is pathological, for the following reason. The saddle point solutions that give rise to density of states at $E=0$ involve potential fluctuations of magnitude $W \chi_0 A^2 \approx W \frac{1}{W \xi A^2 } A^2 \approx 1/\xi$. In the white noise limit the bound states require increasingly large potential fluctuations. If we work with a model of unbounded disorder (such as the model used in this section) then we obtain an order one density of states in the limit $\xi \rightarrow 0$. However, in this limit the solutions are singular at $r \rightarrow 0$, and require  potential fluctuations of diverging magnitude. If the disorder fluctuations are ultimately bounded, then a different analysis is required. For disorder fluctuations that are bounded by $c \Lambda$, the analysis in the present subsection applies for $\xi > a/c$. Meanwhile, the weak disorder limit is $W^2/\xi \ll 1$, or equivalently $\mu_0 b \ll 1$.

\section{Transport near the Dirty Dirac point}
Thus we have shown that (notwithstanding perturbative arguments to the contrary), the density of states at a disordered 3D Dirac point does not vanish, even for arbitrarily weak disorder. We now turn our attention to the transport properties. A systematic approach to transport properties would involve writing down a supersymmetric sigma model (or a replica sigma model), and incorporating the effect of the localized saddle point solutions identified in Section IV. (For a discussion of how to translate a saddle point calculation of the form developed in Section IV to the supersymmetric and replica formalisms, see \cite{Yaida}). However, emboldened by the success of our scaling arguments in calculating the density of states, we now choose instead the simpler and more intuitive option of generalizing our scaling arguments to a scaling theory of transport near the 3D Dirac point. That will be the focus of this section.

We note that while traditional Lifshitz tails involve exponentially bound states that live in a band gap, here we are dealing with power law bound resonances that co-exist with a continuum of extended states (albeit a continuum that has vanishing density of states). The resulting transport behavior will be very different to that encountered with traditional Lifshitz tails.

We recall that a single rare potential well with width $b$ and depth $\lambda$ has a cross section for states at an energy $E$ that takes the form
\begin{eqnarray}
\sigma(E, \lambda) &\sim& \frac{E^2 b^2 }{(\lambda - \lambda_c(E))^2 + E^4 b^2}; \nonumber\\  \lambda_c(E) - \lambda_c(0) &\sim& E; \qquad \lambda_c(0) \sim \pm \pi / b
\end{eqnarray}
i.e. there is a line of resonances in the $(\lambda, E)$ plane, with width $\delta \lambda \sim \delta E \sim b E^2$. It is instructive to calculate the mean free path from scattering off resonant rare regions. This behaves as
\begin{equation}
l \approx \frac{1}{\int d \lambda P(\lambda) \sigma(\lambda, E)} \sim (\nu_0 b) ^{-1} ~.
\end{equation}
where $\nu_0$ is given by (\ref{dos1}) and we recall that we are working with a model of disorder where $P(\lambda) \sim \exp(- \lambda^2 / 2 \mu_0^2)$.

At high energy where the SCBA remains valid, the resulting mean free path is $l\sim 1/(\mu_0^2 b^3 E^2)$.  The rare regions start to dominate the scattering when this SCBA mean free path exceeds that due to the rare regions, which is at an energy scale $ E \lesssim \sqrt{\nu_0}/(\mu_0 b)$. However, rare regions do not start to dominate the density of states until $E \lesssim \nu_0^{1/2}$ (which is a much smaller energy scale, in the weak disorder limit $\mu_0 b \ll 1$).  Moreover, we do not enter the strong scattering / hopping conduction regime until $E < \nu_0 b$ (according to the Ioffe-Regel criterion \cite{IoffeRegel}). Thus, we are led to identify four distinct regimes. At the highest energies $E \ge \sqrt{\nu_0}/\mu_0 b$, the behavior is governed by SCBA.  For $\nu_0^{1/2} < E < \nu_0^{1/2}/\mu_0 b$, the DOS is dominated by extended states, but the (still weak) scattering is dominated by rare regions. Meanwhile, in the regime $\nu_0 b < E < \nu_0^{1/2}$, the DOS and scattering are dominated by the rare regions, but the mean free path is still much longer than the wavelength and the scattering is in this sense weak. Finally, for $E < \nu_0 b $, the mean free path is less than $1/E$, and we are in the `strong scattering' regime where it no longer
makes sense to talk about weakly-scattered extended states.  In this regime, the states all live on rare regions, and transport proceeds by hopping.
In this region we have $\delta \lambda \sim \delta E \sim b E^2 \sim b^3 \nu_0^2$, and the typical hopping rate is also $b^3 \nu_0^2$.
Meanwhile, the density of rare regions is $P(\lambda_c) \delta \lambda \sim b^3 \nu_0^3$, and the typical spacing is $(b \nu_0)^{-1}$.
Thus, transport in this regime occurs due to hopping over length scales $(b \nu_0)^{-1}$.%, consistent with the result quoted in the main text. %We note too that for $\delta \lambda \sim b^2 \nu_0^2$, the `constant' piece in the bound state wavefunctions is of the same order as the $1/r^2$ piece at a distance $(b \nu_0)^{-1}$.

In both intermediate energy regimes, the carriers spend a typical time $\sim b^{-1} E^{-2}$ trapped on each resonant special well (this is just the width of the resonance) whereas the time spent traveling
freely in between special wells is proportional to the mean free path $l \sim (b\nu_0)^{-1}$. %As $E \rightarrow \sqrt{\nu_0}$, the time spent traveling freely is of the same order as the time spent trapped on resonances, but for
Thus, in the intermediate energy regime $\nu_0 b < E < \nu_0^{1/2}$, the time spent trapped on resonances is much longer than the time spent traveling freely,
whereas in the intermediate energy regime $\sqrt{\nu_0} < E < \sqrt{\nu_0 b/W^2}$, the time spent traveling freely exceeds the time spent trapped on resonances.

Some properties of each of our four regimes are summarized in Table I.  In each case the diffusivity is $D\sim l^2/\tau$, with $l$ the typical hopping distance in the hopping regime and the mean free path in the other regimes.
The time between hops or scattering events is $\tau$.  The  zero-temperature conductivity for these noninteracting carriers is then
$\sigma_{DC} =  \nu e^2 D$, where $\nu$ is the DOS.
Stitching together the low energy (hopping dominated) and high energy (SCBA) regions leads to the plot Fig.1. %The various scaling regimes (and their properties) are summarized in Table 1.

\begin{table*}
\begin{tabular}{|c|c|c|c|c|c|c|}
\hline
Energy regime & Description & Length scale & Time scale & DOS & Diffusivity & DC conductivity\tabularnewline
\hline
$E <  (\hbar v)^2 \nu_0 b$ & Hopping & $(\hbar v \nu_0 b)^{-1}$ &  $(\hbar^2 v^3 \nu_0^2 b^3)^{-1}$& $ N \nu_0 $& $v b $ & $N e^2 \nu_0 v b$\tabularnewline
\hline
$(\hbar v)^2 \nu_0 b < E < (\hbar v)^{3/2} \nu_0^{1/2} $ & Intermediate I  & $(\hbar v \nu_0 b)^{-1}$ & $\frac{\hbar^2 v}{E^2 b}$ & $ N \nu_0 $& $\frac{E^2}{\hbar^4 v^3 b \nu_0^2 } $ & $\frac{Ne^2  E^2}{ \hbar^4 v^3 b \nu_0 } $\tabularnewline
\hline
$(\hbar v)^{3/2} \nu_0^{1/2} < E < (\hbar v)^{5/2} \nu_0^{1/2}/\mu_0 b $ & Intermediate II  & $(\hbar v \nu_0 b)^{-1}$ & $(\hbar v^2 \nu_0 b)^{-1}$ & $ N \frac{E^2}{(\hbar v)^3} $& $\frac{1}{\hbar b \nu_0}$ & $\frac{Ne^2 E^2}{ \hbar^4 v^3 b \nu_0} $\tabularnewline
\hline
$(\hbar v)^{5/2} \nu_0^{1/2}/\mu_0 b < E$ & SCBA & $\frac{(\hbar v)^4}{\mu_0^2 b^3 E^2}$ & $\frac{\hbar^4 v^3}{\mu_0^2 b^3 E^2}$ & $ N \frac{E^2 }{(\hbar v)^3}$& $\frac{\hbar^4 v^5}{\mu_0^2 b^3 E^2} $ & $N \frac{e^2}{\hbar} \frac{(\hbar v)^2}{\mu_0^2 b^3}$ \tabularnewline
\hline
\end{tabular}
\caption{Table listing the scaling properties of the four distinct energy regimes (up to purely numerical prefactors). Here $N$ is the number of Dirac points,
 and $\nu_0$ is the (exponentially small) zero energy density of states per unit volume. We have explicitly displayed factors of $\hbar$ and $v$ for clarity, although the discussion in the main text is in terms of natural units $\hbar = v = 1$. 
We have used $W^2 = \mu_0^2 b^3$  to denote the disorder strength, where $b$ is of order the disorder correlation length $\xi$. The results assume we are in the limit of weak disorder, $\hbar^2 v^2 b/W^2 \gg 1$.  The `Length scale' column lists the typical hopping distance
in the hopping regime, and the mean free path in all other regimes. The `Time scale' column lists the typical hopping time in the hopping regime, the
typical dwell time on a resonant well in intermediate regime I, and the scattering time in the other two regimes.
%the ratio of time spent travelling freely to time spent trapped on resonances for a typical electron.
The rest of the columns seem self explanatory.  %In the SCBA regime, resonances make only a subleading contribution to scattering, and the ratio cannot be easily estimated (although it is much greater than one).
For the estimates of the transport in the hopping regime, we assume those states are not localized and the carriers do a random walk with the step length and time set
by these scales; this is what happens in the other regimes.}
\end{table*}

We note that when estimating the diffusion constant we ignore the possibility of interference between distinct paths. Such interference could give rise to localization or anti localization behavior in the hopping model at very long length scales. Now the fact that the wave functions have a $1/r^2$ falloff (which is slower than $1/r^d$) guarantees that there can be no localization on an individual resonance. However, at the very longest length scales we have a theory of non-interacting fermions hopping on a random network of (exponentially widely spaced) resonances, and we may worry about interference between distinct paths on this network. Within the model of purely scalar potential disorder considered here, the various Dirac points are all decoupled. It is widely believed that one cannot localize a single Dirac fermion. This belief is based on calculations involving sigma models, which are generally designed to treat fluctuations about a translation invariant saddle point. We have shown that a translation invariant saddle point is not the appropriate starting point, and thus the sigma models need to be rederived. Assuming that it remains impossible to localize a single Dirac fermion even taking rare region effects into account, the only possibility to be wary of is anti localization. However, the standard scaling arguments suggest that anti localization should be a weak effect in three dimensions (at least for weak disorder $W\rightarrow 0$), with the $\beta$ function for the conductance taking the form $\beta(g) \sim 1 + f(W)$, where $f(W\rightarrow 0) = 0$. Thus we conclude that the neglect of interference between distinct paths is not a real problem, and that transport at the lowest energies should indeed be diffusive, and dominated by hopping between rare resonances.

\begin{figure}
\includegraphics[width =  0.9\columnwidth]{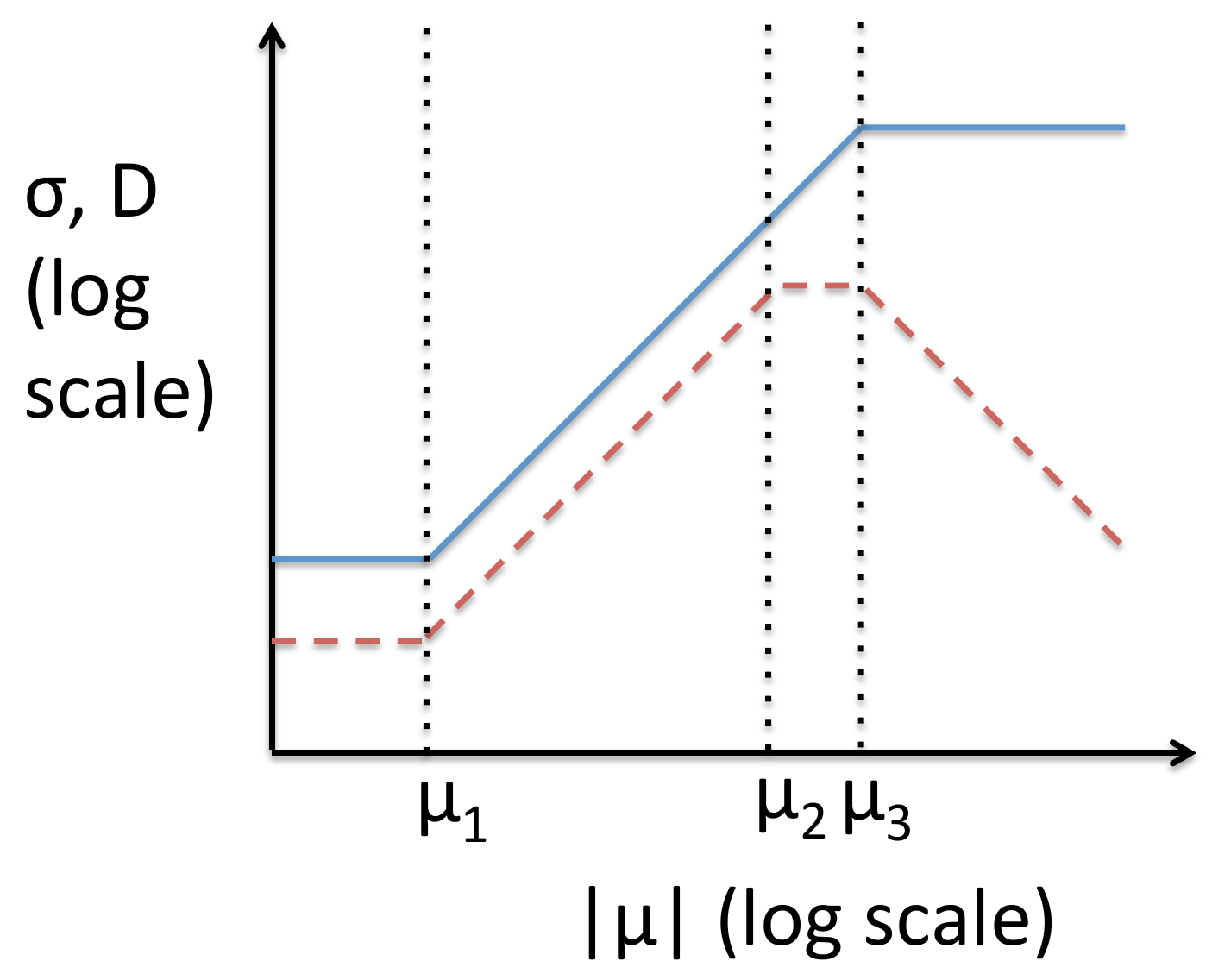}
\caption{\label{fig: conductivity} Schematic behavior of the zero-temperature DC conductivity $\sigma$ (solid blue line) and diffusivity $D$ (dashed red line)
as a function of the chemical potential $\mu$ for disordered non-interacting massless 3D Dirac fermions.  The dotted vertical lines are guides to the eye. Moving from low to high energy, the sequence of regimes and their boundaries is: hopping regime, $\mu_1\sim(\hbar v)^2\nu_0 b$, intermediate regime I, $\mu_2\sim(\hbar v)^{3/2}\nu_0^{1/2}$, intermediate regime II,
$\mu_3\sim(\hbar v)^{5/2}\nu_0 ^{1/2}/\mu_0 b$, SCBA regime.  The density of states is $\nu_0$ in the first two regimes, where it is dominated by rare regions of
linear size $b$.  The rare regions dominate the scattering for all regimes other than the highest-energy SCBA regime.
The nonzero slopes on this log-log plot are $\pm 2$.
%Rare regions dominate the DOS for $|\mu| < \nu_0^{1/2}$, and dominate transport for $|\mu| < (\nu_0 b/V)^{1/2}$. At large doping $|\mu| > (\nu_0 b/V)^{1/2}$, the SCBA applies and conductivity is a constant, while the diffusion constant falls as $1/\mu^2$. At the smallest doping levels $|\mu| < \nu_0$, the conductivity saturates to an exponentially small value $\sigma_{\infty} \nu_0$, while the diffusion constant saturates to $\sigma_{\infty} e^{-2}$. In the intermediate regime $\nu_0 < |\mu| < \nu_0^{1/2}$, the conductivity and diffusion constant both scale as $E^2$, whereas in the intermediate regime $\nu_0^{1/2} < |\mu| < (\nu_0 b/V)^{1/2}$, the diffusion constant is constant, but the conductivity scales as $\sim E^2$.
For more details, see text and Table I}.
\end{figure}

\section{Interplay of disorder and interactions}
Thus far we have concentrated on non-interacting 3D Dirac fermions. We now turn our attention to the interplay of disorder and interactions. Above a critical interaction strength, repulsive interactions destroy the Weyl semimetal phase \cite{Wei, Zhang, Maciejko}. Subcritical repulsive interactions suppress (charged) rare regions, and reduce the rare-region DOS at the Dirac point. We defer further consideration of repulsive interactions to future work.
%However, a much more
Instead, we now consider the interesting interplay that occurs between disorder and {\it attractive} interactions at the Dirac point.

Attractive interactions above a critical strength will trigger superconductivity in the clean system \cite{Meng, Cho}. Subcritical interactions will produce
local pairing on rare regions where the local DOS is non-zero over a larger length scale than the local coherence length $\Xi$. Establishment of phase coherence between
islands by Josephson coupling will then drive the system into a (granular) superconducting state at sufficiently low temperatures. We have discussed similar phenomena for the 2D Dirac system in \cite{DiracSC}. %In the weak coupling limit, superconductivity will arise purely due to rare region effects.
We focus on estimating the energy scale for the superconducting state, in the presence of a random scalar potential that is approximately Gaussian distributed (for small fluctuations), but which is ultimately bounded, with no local fluctuations
%of the chemical potential
that are larger than $\Lambda$.

Local pairing occurs in islands of local average potential $\mu$ and size $L \ge \Xi$, where $\Xi \sim (v/\omega_D) \exp(1/ G \mu^2)$
is the local coherence length in the BCS approximation, $\omega_D$ is the Debye frequency and $G$ is the strength of the attraction in the leading pairing channel. %  {\bf g needs to be defined.  can it be defined so the $4\pi$ in the exponent is meaningful?
Integrating over $L$ in a saddle point approximation, we find the result is dominated by islands of size $L \cong \Xi$.
The probability of finding such an island is
\begin{equation}
P_{SC} \sim  \int_0^{\min(\Lambda,\frac{1}{G^{1/2}})} d \mu \exp\left(- \frac{ \mu^2}{2\mu_0^2 \omega_D^3 b^3} \exp (3 / G \mu^2) \right) ~,
\end{equation}
where %$\Lambda$ is the bandwidth, and 
$  G^{-1/2}$ marks the boundary of the weak coupling BCS regime.
This is dominated by the regions near the upper limit of this integral, %close to the cutoff, 
and yields %Evaluated in a saddle point approximation, the probability of finding superconducting islands is
%island
\begin{eqnarray}
P_{SC}(G) &\sim& \exp \left( - \frac{  f(G)}{\omega_D^3 R^3} \right) ~,   \label{eq: pasc} \\
f\big(G<G_1 \big) &\sim& \frac{\Lambda^2}{\mu_0^2} \exp\big(\frac{3 }{G \Lambda^{2}}\big) ~, \quad f\big(G_1 < G \ll G_c\big) \sim \frac{1}{G \mu_0^2}. \nonumber
\end{eqnarray}
Here $G_1 = \frac{1}{\Lambda^{2}}$, and $G_c$ is the critical coupling for superconductivity in the clean system. %, and $f$ is correct upto numerical pre-factors.
This density of superconducting islands is doubly exponentially small in $G$ for $G \rightarrow 0$ when even the maximally doped islands with local
$\mu \approx \Lambda$ have to be exponentially large;
but is only exponentially small in $G$ for intermediate $G$, when small superconducting islands with local doping $\mu \le \Lambda$ can form. %The max function ensures that islands have a minimum size $R$ for any $g$.
%Meanwhile, recall that the probability of having a `live' island scaled as $\nu_0^{3/2} \sim \exp(-3l_0 v^2/2 \mu_0^2)$. Thus we see that the probability of having a superconducting island scales the same way with $\mu_0$ as the probability of having a Griffiths region. However, we have an additional parameter $g$ to play with. We will assume that $g$ is sufficiently small such that $P_{SC} < \nu_0^{3/2}$, otherwise the analysis makes no sense.

In the intermediate range of $G$, the energy scale for local Cooper pairing in each of the dominant islands is of order $\hbar\omega_D$. However, the sample will exhibit global
superconductivity only if phase coherence is established between islands. The Josephson coupling between distant islands $J$ may be determined by generalizing the
calculation in \cite{Gonzalez} to the 3D Dirac point. We find that $J \sim 1 /r^5$.
Since the Josephson coupling falls off with distance faster than $1/r^3$, the coupling between nearest neighbor islands dominates. The system of locally superconducting islands embedded in a semimetal then establishes global phase coherence on temperature scales smaller than the typical nearest neighbor Josephson coupling. This leads to an estimated critical temperature for phase ordering
\begin{equation}
T_{c} \sim \omega_D / r^5 \sim \omega_D P_{SC}^{5/3} \sim \omega_D \exp \left( - \frac{5 f(G)}{3  R^3 \omega_D^3 } \right) ~. \label{eq: tcunbounded}
\end{equation}

We have implicitly assumed that the pairing is $s$-wave. If the `local pairing' was not $s$-wave, then the Josephson couplings would be frustrated, and the ground state would be a `gauge glass' \cite{glass}. We leave further discussion of non-$s$-wave orders to future work, noting only that in \cite{Cho} it was determined that $\delta$-function attraction in the clean system favors $s$-wave pairing.

\subsection{ Attractive and repulsive interactions:} We now discuss the situation when Coulomb repulsion coexists with retarded attractive interactions. We assume that the Morel-Anderson condition \cite{Anderson-Morel} is satisfied, so that local pairing on islands still occurs. However, the effective Hamiltonian for the islands must now contain not only the Josephson couplings, but also charging effects (electrostatic interactions may be neglected due to screening \cite{screening}). Thus, the effective Hamiltonian for the islands is
\begin{equation}
H = \sum_{i}  (E_c  n^2_i + V_i n_i) %+ \sum_{ij} \frac{e^2}{r_{ij}} n_i n_j
+ \sum_{\langle ij \rangle} J_{ij} \cos(\phi_i - \phi_j) ~,
\end{equation}
where $i$ and $j$ label superconducting islands, $\phi_i$ is the phase of the $i^{th}$ island, and $n_i = i \partial/\partial \phi_i.$
The Josephson couplings $J_{ij}$ operate primarily between nearest-neighbor islands, as previously discussed, and the $V_i n_i$ term reflects the random
scalar potential on the islands.
Such Hamiltonians have been long discussed in the theory literature \cite{Fisher, Svitsunov, Altman}, and are known to support %three distinct phases.
%In addition to the superconducting phase (large $J/E_c$), and the phase with `Mott insulating' islands embedded in the Weyl semimetal (small $J/E_c$),
a superconducting phase, and also a Bose glass \cite{Mott}. The glassy phase is characterised by an infinite superconducting susceptibility, but no long range order,
and has a regime of stability that grows larger as the system becomes more disordered.% (larger variations in $J_{ij}$ and $V_i$).

\section{Conclusions}
Thus we have demonstrated that a $3D$ Dirac point has a non-vanishing density of states $\nu_0 \sim \exp(-\xi/W^2)$ for weak scalar potential disorder with strength $W$ and correlation length $\xi$. The physics at low energies is dominated by exponentially rare, power law bound resonances which break translation symmetry and `pull' density of states down to zero energy. We have shown how the density of states can be estimated using a rare regions scaling argument, and also using a systematic saddle point analysis. The systematic saddle point analysis also reveals what was missed by the existing theoretical arguments for the irrelevance of disorder (detailed in Section II): those arguments only considered fluctuations about a translation invariant saddle point, whereas the non-zero density of states arises due to other, translation non-invariant saddle points, which cannot be accessed through perturbation theory about a translation invariant saddle point.

We have also constructed a scaling theory of transport near a 3D Dirac point in the presence of random scalar potential disorder with strength $W$ and correlation length $\xi$. This theory reveals that there are four distinct transport regimes. At the highest energies $|E| > \sqrt{\nu_0 \xi}/{W}$, the SCBA solution applies and both scattering and the DOS are dominated by extended states. The DOS scales as $E^2$ and the DC conductivity is constant. For $\sqrt{\nu_0} < |E| < \sqrt{\nu_0\xi }/W$, scattering is dominated by rare regions, but the DOS is still dominated by extended states. In this regime the DOS and the DC conductivity both scale as $E^2$. For $\nu_0 \xi< |E| < \sqrt{\nu_0}$ both scattering and the DOS are dominated by rare regions. The DOS is constant, but the conductivity scales as $E^2$. Finally, for $|E| < \nu_0 \xi$, we enter a `strong scattering' regime in which we argue that both the DOS and DC conductivity saturate to non-zero constants.

Finally, we have also discussed the interplay of attractive interactions with rare resonances, which can drive the system into a granular superconducting phase, with a critical temperature that we estimate. We have also discussed the Bose glass phases that can arise in the presence of both attractive and repulsive interactions.

This work has established that the existing framework for thinking about 3D Dirac points - in terms of translation invariant disorder averaged theories - is inaccurate at the lowest energies. Instead, one must take into account the effects of rare resonances, which control the physics close to the Dirac point. In light of the rapid experimental advances in synthesizing materials supporting 3D Dirac points, we hope that it will soon be possible to probe the asymptotic low energy regime in experiments, and to directly test the scaling theory advanced in this paper.

{\bf Acknowledgements}: We thank S.A. Parameswaran for a useful discussion. This research was supported in part by the National Science Foundation under Grants No. DMR08-19860 (DAH) and DMR 10-06608 (SLS), and by a PCTS fellowship (RN).

 \section{Appendix}
 In this appendix we solve the three dimensional Dirac equation in a spherically symmetric potential. The results obtained in this way will be essential to construction of our argument. The Dirac equation in a spherically symmetric potential can be written as
\begin{equation}
\big( -i \hbar v \sigma_i \partial_i + V(\vec{r}) - E\big) \psi(\vec{r}) = 0
\end{equation}
We work with a single Dirac point, since the different Dirac points are all decoupled. The resulting equation for a two component spinor wave function is sometimes also referred to as the Weyl equation. However, we continue to refer to it as a Dirac equation here, to emphasize that our results are not particular to Weyl semimetals. 

The eigenstates of the Dirac Hamiltonian are also eigenstates of total angular momentum $j$, but are not eigenstates of orbital angular momentum $l$. Using the standard Pauli matrix multiplication identity $\sigma_i \sigma_j = \delta_{ij} + i \epsilon_{ijk} \sigma_k$, we rewrite the gradient term as
\begin{eqnarray}
\sigma_i \partial_i &=& \frac{\sigma_i r_i}{r_j r_j} \sigma_k r_k \sigma_l \partial_l = \frac{\vec{\sigma}\cdot \hat{r}}{r}(r_l \partial_l + i \epsilon_{klm} r_k \partial_l \sigma_m)\nonumber\\ &=& \frac{\vec{\sigma}\cdot \hat{r}}{r}\big(r \frac{\partial}{\partial r} + i \vec{\sigma} \cdot ( \vec{r} \times \vec{\partial})\big) = \vec{\sigma}\cdot \hat{r} \big(\partial_r - \frac{\vec{\sigma} \cdot \vec{L}}{\hbar r}\big)\nonumber
\end{eqnarray}
using the notation $\hat{r} = \vec{r}/r$ and $r^2 = r_j r_j$, and where $\vec{L}$ is the usual quantum mechanical angular momentum operator. This prompts us to search for a solution of the form $\psi = R(r) \phi$, where $R$ is a scalar function that depends purely on radius, whereas $\phi$ is a two component spinor which is an eigenstate of the angular momentum operator, and which is independent of radius.

Now, the eigenstates of the operator $\vec{\sigma} \cdot \vec{L}$ are two component spinors $\phi^{\pm}_{j,j_z}$ with total angular momentum $j$, angular momentum projection onto the z-axis $j_z$, and orbital angular momentum $l_{\pm} = j \mp 1/2$, which take the explicit form \cite{Callan}
\begin{equation}
\phi_{j, j_z}^{\pm}  = \left( \begin{array}{c} \sqrt{\frac{l_{\pm}+1/2 \pm j_z}{2l+1}} Y^{l_{\pm}}_{j_z-1/2} \\ \pm \sqrt{\frac{l_{\pm}+1/2 \mp j_z}{2l_{\pm}+1}} Y^{l_{\pm}}_{j_z+1/2} \end{array} \right)
\end{equation}
where the $Y$ functions are the usual spherical harmonics. We note that the $\pm$ superscript refers to the angular structure. Using the identities $\vec{J} = \vec{L} + \frac12 \vec{\sigma}$ and $\vec{J}\cdot \vec{J} = j(j+1) \hbar^2 $, $\vec{L} \cdot \vec{L} = l(l+1) \hbar^2$, we can show that the spinors obey $\vec{\sigma} \cdot \vec{L} \phi^{\pm}_{j, j_z} = - (1+ \kappa)\hbar  \phi^{\pm}_{j, j_z}$, where $\kappa = -(j+1/2)$ is a negative integer for $\phi^+$ and $\kappa = j+1/2$ is a positive integer for $\phi^-$.

We note that the functions $\phi_{j, j_z}^{\pm}$ have orbital angular momentum differing by one, and thus have opposite parity under inversion. Since $\vec{\sigma} \cdot \hat{r}$ commutes with the angular momentum operator and changes sign under inversion, it follows that it must turn $\phi^+$ into $\phi^-$ and vice versa. Since the gradient term mixes the angular sectors $\phi^{\pm}$, the eigenstates of the Hamiltonian must be linear superpositions of pieces with $\phi^+$ and $\phi^-$ angular structure. Thus we find that the eigenstates in the vicinity of the Dirac point take the form %Translating \cite{Callan} to the Weyl basis, we find that the eigenstates in the vicinity of the Dirac points at $\pm \vec{Q}$ take the form
\begin{equation}
\psi^{\pm} = f(r) \phi^{\pm}_{j, j_z} +i  g(r) \phi^{\mp}_{j, j_z};
\end{equation}
where $f$ and $g$ are purely radial functions with no angular dependence.
 Substituting this expression for the wavefunctions into the Dirac equation leads to the two equations
\begin{equation}
\frac{1}{\hbar v}(E - V) f = \partial_r g + \frac{1- \kappa}{r} g ; \quad -\frac{1}{\hbar v}(E-V)g = \partial_r f + \frac{1+\kappa}{r} f \label{two}
\end{equation}
where $\kappa$ is a positive integer for one solution, and $\kappa$ is a negative integer for its degenerate partner which differs only in its angular structure. Let us pick positive $\kappa$ for specificity.

\subsection{Square wells}

We begin by considering a `square well' potential $V(\vec{r}) = \lambda \Theta(b - r)$, although we will relax this approximation in due course. We note that because of the particle-hole symmetry of the problem positive and negative $V$ must yield identical results. Substituting the square well potential into (\ref{two}) and performing some elementary manipulations then leads to the equation
\begin{equation}
r^2 \partial^2_r f + 2r  \partial_r f +\big( \frac{(\lambda \Theta(b-r)-E)^2 r^2}{\hbar^2 v^2} -  \kappa (1+\kappa)\big)f =0
\end{equation}
We recognize this as the spherical Bessel equation, whose solutions are spherical Bessel functions. Substituting $f$ back into the equation for $g$ then determines $g$. Thus, the solutions for arbitrary $E \neq V$ take the form
\begin{widetext}
\begin{eqnarray}
f(r) = \frac{A}{\sqrt{|V-E|r/\hbar v}}J_{\kappa+1/2}\big(|V-E|r/\hbar v\big) +  \frac{B}{\sqrt{|V-E|r/\hbar v}}K_{\kappa+1/2}\big(|V-E|r/ \hbar v\big) \\
g(r) = \sign(V-E)\left( \frac{A}{\sqrt{|V-E|r/\hbar v}} J_{\kappa - 1/2}(|V-E| r/ \hbar v) +  \frac{B}{\sqrt{|V-E|r/ \hbar v}}K_{\kappa-1/2}\big(|V-E|r/ \hbar v\big) \right)
\end{eqnarray}
\end{widetext}
where $J$ and $K$ are Bessel functions of the first and second kind respectively. To save writing, we now adopt a system of units where $\hbar v = 1$. We will re-introduce $\hbar v$ whenever necessary for clarity.

 For $r<b$, $V= \lambda$. In this region, we must have $B=0$ to have a regular solution at the origin. Meanwhile, for $r>b$, $V = 0$. In this region we can have $A' \neq 0$ and $B' \neq 0$. Thus, we have
 \begin{widetext}
\begin{eqnarray}
f(r) = \frac{A}{\sqrt{|\lambda-E|r}}J_{\kappa+1/2}\big(|\lambda-E|r\big) \Theta(b-r) + \left(\frac{A'}{\sqrt{|E|r}}J_{\kappa+1/2}\big(|E|r\big) +  \frac{B'}{\sqrt{|E|r}}K_{\kappa+1/2}\big(|E|r\big) \right) \Theta(r-b)\nonumber \\
g(r) = \sign(\lambda-E) \frac{A}{\sqrt{|\lambda-E|r}} J_{\kappa - 1/2}(|\lambda-E| r) \Theta(b-r) - \sign(E) \left( \frac{A'}{\sqrt{|E|r}} J_{\kappa - 1/2}(|E| r) +  \frac{B'}{\sqrt{|E|r}}K_{\kappa-1/2}\big(|E|r\big) \right)\Theta(r-b)\nonumber \end{eqnarray}
Since we are dealing with a first order differential equation, only the wave function need be continuous (there is no requirement that derivatives be continuous). Imposing continuity of the wave function then implies that
\begin{equation}
\left(\begin{array}{c} A' \\ B' \end{array}\right) = A \sqrt{|E|/|\lambda-E|} \frac{1}{\Delta} \left(\begin{array}{cc} K_{\kappa - 1/2}(|E|b) & -K_{\kappa + 1/2}(|E|b) \\ - J_{\kappa - 1/2}(|E|b) & J_{\kappa+1/2}(|E|b) \end{array} \right)\left( \begin{array}{c} J_{\kappa + 1/2}(|\lambda - E|b)  \\ \sign(\frac{E}{E-\lambda}) J_{\kappa - 1/2}(|\lambda - E|b)\end{array}\right)
\end{equation}
\end{widetext}
where $\Delta$ is the determinant of the $2\times2$ matrix. This fails for special values of $E$ where the matrix is singular (vanishing determinant).

We note that continuity of the wave function also implies continuity of the probability density (given by the norm squared of the wave function). The norm squared of the wave function at $r=b$ (defined as $|f|^2 + |g|^2$) never vanishes, and scales as $(\lambda - E)^{-2} b^{-2}$ in the limit of large $|\lambda - E|b$ while saturating to a constant in the limit of small $|\lambda - E|b$. %We can identify Fig.\ref{fig: normsquare}a as the probability density just outside the well.
Thus, the probability density just outside the well never vanishes,
%
%\begin{figure}
%a)\includegraphics[width = 0.45\columnwidth]{normsquared}
%b)\includegraphics[width = 0.45\columnwidth]{normsquare}
%\caption{\label{fig: normsquare} Plot showing (a) $|\psi(r=1)|^2$ (probability density) as a function of $\lambda - E$ for $\kappa = 1$ and (b) $|\psi(r=1)|^2 (\lambda-E)^2$ as a function of $\lambda - E$. We see that there is always substantial leakage of probability density out of the well.}
%\end{figure}
%
and there is always `leakage' of the probability density out of the region $r<b$. Moreover, the spherical Bessel functions only decay as $1/r$ at long distances, so the probability density only decays as $1/r^2$ at long distances. Thus, the solutions constructed above are not normalizable in an infinite volume.

A qualitatively different (and properly normalizable) exterior solution exists when $E = 0$. When $E=0$ then the two equations in (\ref{two}) decouple for $r > b$, and can be straightforwardly solved to give an exterior solution
\begin{eqnarray}
f(r>b) &\sim& r^{-(1+\kappa)} \textrm{ or } f(r) = 0; \nonumber\\
g(r>b) &\sim& r^{\kappa - 1} \textrm{ or } g(r) = 0
\end{eqnarray}
Recall that $\kappa$ is a positive integer. This corresponds to a bound state if and only if we pick the solution $g(r) = 0$, which comes about if $g(r)$ is matched to a node of the interior Bessel function. This in turn happens only for special values of the well depth $\lambda_c$. The probability density in this bound state decays like $1/r^4$ outside the well (i.e. most of the probability density is localized on the well and the solution is properly normalizable).

Although there is a well depth corresponding to a bound state for all values of $\kappa$, larger values of $\kappa$ require a deeper (or wider) well in order to have a bound state. The physics of interest to us will thus be controlled by the minimal well, which has a bound state for $\kappa = 1$. Bound states with $\kappa = 1$ arise when  $\lambda b  \approx m \pi $, where $m$ is a positive integer. Again, values of $m$ greater than one involve deeper or wider wells, and the minimal well which controls the physics has $\kappa = 1$ and $m=1$, with a well depth $\lambda_c \approx \pi / b$. Note that there is a single parameter that must be tuned to get a bound state: either we can fix $b$ and tune $\lambda$, or we can fix $\lambda$ and tune $b$.

We note that the angular eigenfunction $\phi_-$ has total angular momentum $j = 1/2$ (for $\kappa = 1$), but may have $j^z = \pm j$. Thus there are two bound states corresponding to the $\kappa = 1$ solution identified above. We note that there are two additional bound states corresponding to $\kappa = -1$ and $\lambda = \lambda_c = \pi/b$, which now corresponds to an angular eigenfunction $\phi_+$ and has $f(r) = 0$. Thus, there are four bound states per Dirac point for each special well.

\subsection{Beyond square wells}
Thus far we considered square well potentials. Now we consider a potential that has a long range tail. For specificity, we consider the potential $V(r) = \lambda \Theta(b-r) + \epsilon(r) \Theta(r-b)$, where $\epsilon(r) = \lambda b^4/r^4$. The equations (\ref{two}) for zero energy states in the domain $r> b$ then become
\begin{equation}
\frac{\lambda b^4}{r^4}f(r) = \partial_r  g(r); \qquad -\frac{\lambda b^4}{r^4}g(r) = \left(\partial_r + \frac{2}{r} \right) f(r).
\end{equation}
some elementary manipulations allow us to rewrite this as a single equation for $g$, which takes the form
\begin{equation}
r^8 \partial_r^2 g(r) + 6 r^7 \partial_r g(r) + \lambda^2 b^8 g(r) = 0
\end{equation}
This differential equation can be solved on {\it Mathematica}, and has the analytic solution
\begin{equation}
g(r) = \frac{C_1 (\lambda b^4)^{5/6}}{r^{5/2}} J_{-5/6}(\lambda b^4/3r^3) + \frac{C_2 (\lambda b^4)^{5/6}}{r^{5/2}} J_{5/6}(\lambda b^4/3r^3)
\end{equation}
In the $r\rightarrow \infty$ limit, the first term asymptotes to a constant, while the second term falls off as $1/r^5$. Since we want a bound state solution, we set $C_1=0$ and thus obtain the solution
\begin{widetext}
\begin{eqnarray}
f(r>b) &=& C \frac{V_0^{5/6} J_{11/6}(V_0/3r^3) - 5 V_0^{-1/6} r^3 J_{5/6}(V_0/3r^3) -  V_0^{5/6}J_{-1/6}(V_0/3r^3)}{2r^{5/2}};\nonumber\\
g(r>b) &=& \frac{C V_0^{5/6}}{r^{5/2}} J_{5/6}(V_0/3r^3) %\qquad g(r\rightarrow \infty) \sim 1/r^5.
% \\ f(r \rightarrow \infty) &\sim& -\frac{V_0^{2/3}(5\Gamma(5/6) + 6 \Gamma(11/6))}{2(6^{5/6} \Gamma(5/6) \Gamma(11/6))x^2}\sim -C \frac{V_0^{2/3}}{r^2} + O(1/r^8)
\end{eqnarray}
\end{widetext}
where we have defined the shorthand $V_0 = \lambda b^4$. In the limit $r \rightarrow \infty$, this has the asymptotics $f(r) \sim 1/r^2 $ and $g(r) \sim 1/r^5$, i.e. at long distances the probability density decays as $1/r^4$ (a properly normalizable behavior). However, this exterior solution is a proper solution of the Dirac equation only if it can be matched onto the interior solution for $r < b$, which consists of spherical Bessel functions of the first kind, and takes the form.
\begin{eqnarray}
f(r) &=& \frac{A}{\sqrt{|\lambda-E|r}}J_{\kappa+1/2}\big(|\lambda-E|r\big) \nonumber \\
g(r) &=& \sign(\lambda-E) \frac{A}{\sqrt{|\lambda-E|r}} J_{\kappa - 1/2}(|\lambda-E| r) \Theta(b-r)\nonumber \end{eqnarray}
\begin{figure}
\includegraphics[width = \columnwidth]{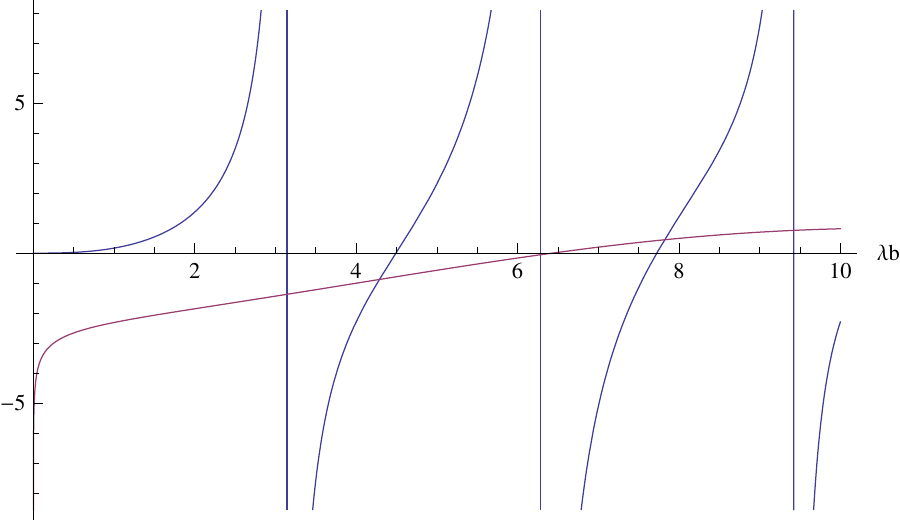}
\caption{\label{graphical solution} The above graph plots the left hand side and right hand side of (\ref{consistency}), as a function of $\lambda b$. The intersections represent values of $\lambda b$ for which a properly normalized solution exists. The intersections of interest to us are the ones where the red line crosses a non-vertical blue line. The intersections with the vertical blue lines involve $A = C = 0$ and thus only give rise to the trivial solution $\psi = 0$}
\end{figure}
Matching requires that $A/C = (\lambda b)^{4/3} J_{5/6}(\lambda b/3)/J_{1/2}(\lambda b) $ and also
\begin{widetext}
\begin{equation}
2 (\lambda b)^{5/6} J_{3/2}(\lambda b) J_{5/6}(\lambda b/3) / J_{1/2}(\lambda b)= J_{11/6}(\lambda b/3) - \frac{5}{\lambda b} J_{5/6} (\lambda b/3) - J_{-1/6}(\lambda b/3) \label{consistency}
\end{equation}
\end{widetext}
Clearly there is a single parameter that can be tuned, namely $\lambda b$. It can be seen graphically (see Fig. \ref{graphical solution}) that the above equation has solutions for particular values of $\lambda b$. Thus, bound states can be obtained by tuning $\lambda b$, just as for the square well, although the critical values for $\lambda b$ are of course different.

It can be readily checked by solving the radial equations numerically on {\it Mathematica}  that the $1/r^4$ potential is not special. Bound states arise also for exponential tails, and for power law tails where the potential falls off faster than $1/r^4$. In all cases obtaining a properly continuous  and normalizable bound state solution requires tuning a single parameter $\lambda b$.
One way to see that there is a single parameter which has to be tuned is the following: the interior solutions have an overall scale factor $A$. The exterior solutions have the overall scale factor $C$. Matching $g$ fixes the ratio of scale factors $A/C$, but we still have to match $f$. Matching $f$ requires tuning one parameter, and the relevant parameter here is $\lambda_0 b$. Moreover, in all cases the probability density decays as $1/r^4$ at large distances, just as for the square well i.e. the asymptotic behavior is unchanged. %It can also be readily determined numerically that the $1/r^4$ tail is not special. 

Thus, we have demonstrated that the square well potential is not special and that qualitatively similar behavior arises for potentials that have a long range tail. However, the square well potential is uniquely convenient for analytical work, and we will use it extensively in the main text.


\begin{thebibliography}{99}
\bibitem[Mott (2013)]{Mott}
With bounded disorder in $V_i$, the large $E_c$ limit may also support a phase with `Mott insulating' islands embedded in the semimetal.
\end{thebibliography}
\end{document}